\documentclass[a4,11pt]{article}
\usepackage{epsfig}

\usepackage{amssymb,epsfig,graphicx,epsf}

\usepackage{amsfonts}
\usepackage{amsmath}
\usepackage{amssymb}
\newcommand{\be}{\begin{equation}}
\newcommand{\ee}{\end{equation}}

\newcommand{\ba}{\begin{eqnarray}}
\newcommand{\ea}{\end{eqnarray}}

\begin{document}

\title{Gravity as an effective theory
\thanks{Presented at the 49th Zakopane School on Theoretical Physics}}

\author{Dom\`enec Espriu\footnote{On leave of absence from October 2009 
at CERN, 1211 Geneva 23, Switzerland}
 and Daniel Puigdom\`enech\\
Departament d'Estructura i Constituents de la Mat\`eria\\ and\\ 
Institut de Ci\`encies del Cosmos (ICCUB)\\ Universitat de Barcelona\\
Mart\'\i ~i Franqu\'es, 1, 08028 Barcelona, Spain.}

\maketitle

\begin{abstract}
Using as inspiration the well known chiral effective lagrangian describing the
interactions of pions at low energies,
in these lectures we review the quantization procedure of Einstein gravity in 
the spirit of effective field theories. As has been emphasized by several authors,
quantum corrections to observables in gravity are, by naive power counting, very small. While
some quantities are not predictable (they require local counterterms of higher 
dimensionality) others, non local, are. A notable example is 
the calculation of quantum corrections
to Newton's law. Albeit tiny these corrections are of considerable theoretical importance,
perhaps providing information on the ultaviolet properties of gravity. We then  try to 
search for a situation where these non local corrections may be observable in a cosmological
context in the early universe. Having seen that gravity admits an effective treatment 
similar to the one of pions, we pursue this analogy and propose a two-dimensional toy 
model where a dynamical zwei-bein is generated from a theory without any metric at all.

\end{abstract}

\vfill
\noindent
October 2009

\noindent
UB-ECM-FP-28/09

\noindent
ICCUB-09-239

\section{Introduction and outline}

This paper summarizes the contents of a set of lectures that were delivered in the 49th
Zakopane School on Theoretical Physics on the subject of treating Einstein theory 
of gravitation as an effective theory and the testable consequences of this procedure, 
and the possibility that gravitons emerge as
Goldstone states after some sort of symmetry breaking mechanism. 
The contents can be basically divided into two parts.
The first one describes the treatment of effective theories taking the chiral lagrangian
of strong interactions as a starting point and proceeding to study 
the gravity case in parallel to the way one 
sets out to quantize the pion lagrangian. This part is not original and we have freely
drawn material from the works of Donoghue\cite{don}, Bjerrum-Bohr\cite{bohr} 
and Khriplovich\cite{khri} in particular.   

The second part contains original work made in collaboration with 
J. Alfaro\cite{aep}, J.A. Cabrer\cite{ce}, T. Multam\"aki\cite{emv} and E. Vagenas\cite{emv}. Some
results are presented in published form for the first time here. In the lectures the subject of 
explicit Lorentz breaking was treated briefly, including some potential 
applications to astroparticle physics\cite{aegs}, but this part is omitted in these written
notes for the sake of homogeneity and consistency of the presentation

We shall start with a succint presentation of the pion chiral lagrangian and the chiral
counting rules. We shall move to the gravity case after that, proceeding to quantize the theory.
An analogous power counting can be implemented in this case too. The power counting turns out to 
be more subtle when matter fields are present, as we shall see. 

Next we will argue why non-local effects are necessarily present and, in fact, that they
provide the only unique and non-ambiguos predictions of quantum gravity at the one-loop level.
These predictions are finite and contribute in a distinctively different way to physical 
observables. This shall be exemplified by studying the first quantum corrections to 
Newton's law and also by analyzing how these corrections may affect the evolution of
a de Sitter universe (inflation). 

Finally, we shall give some credence to the idea that gravitons might be Goldstone bosons 
of some broken symmetry. We are certainly not the first to entertain this idea\cite{ru}, which, on 
the other hand may seem hopelessly flawed due to some in-principle long-standing restrictions\cite{ww}. 
We shall provide a two-dimensional toy model (that, however, can be easily extended 
to four dimensions) that shows that such a mechanism is possible in a model that very much 
parallels the mechanism of chiral symmetry breaking in QCD, and how the theoretical objections
might be circumvented.

\section{Chiral effective theory and chiral counting}

The chiral lagrangian is a non-renormalizable theory describing accurately pion
physics at low energies. It has a long story, with the first formal studies concerning 
renormalizability being due mostly to Weinberg\cite{wein} and later considerably extended by
Gasser and Leutwyler\cite{gl}. The chiral lagrangian contains a (infinite) number
of operators organized according to the number
of derivatives 
\be
{\cal L}= f_\pi^2 {\rm Tr\,} \partial_\mu U \partial^\mu U^\dagger
+ \alpha_1 {\rm Tr\,}\partial_\mu U \partial^\mu U^\dagger \partial_\nu U 
\partial^\nu U^\dagger + \alpha_ 2 {\rm Tr\,} \partial_\mu U 
\partial_\nu U^\dagger \partial^\mu U \partial^\nu U^\dagger + \ldots
\ee
\be 
{\cal L} = {\cal O}(p^2) + {\cal O}(p^4) + {\cal O}(p^6) + ...
\ee
\be 
U\equiv  \exp i \tilde \pi/ f_\pi \qquad \tilde\pi \equiv \pi^a \tau^a/2 
\ee
Pions are the Goldstone bosons associated to the (global) symmetry breaking
pattern of QCD
\be
SU(2)_L \times SU(2)_R \to SU(2)_V
\ee
The above lagrangian is the most general one compatible with the symmetries 
of QCD and their breaking.
Locality, symmetry and relevance (in the renormalization group sense)  are the only 
guiding principles to construct ${\cal L}$. Renormalizability is not. In fact if
we cut-off the derivarive expansion at a given order the theory 
requires countreterms beyond that order no matter how large.

Note that, although the symmetry has been spontaneously broken, the effective 
lagrangian still has the {\it full} symmetry
\be
U\to L U R^\dagger
\ee
i.e. the underlying symmetry is not lost in spite of the (partial) breaking.

Next let us see how a simple power counting in derivatives can be established
at the level of quantum corrections. Let $A_{N^\pi}$ be the amplitude for the
scattering of $N^\pi$ pions. At lowest order in the derivative expansion it will
be of the form
\be
A_{N^\pi}\sim \frac{p^2}{f_\pi^2},
\ee
where $p^2$ represents a generic kinematic invariant constructed with external momenta. At the next order
\be
A_{N^\pi}(p_i)\sim \int \frac{d^4 k}{(2\pi)^4}\, (\frac{1}{f_\pi})^{N^\pi}  
(k^2)^{N_V}
(\frac{1}{k^2})^{N_P},
\ee
where 
$N_V$ and $N_P$ are the number of vertices and propagators, respectively.
Consider e.g. $  \pi\pi \to \pi \pi$ scattering. Then $N^\pi= 4$, $N_V= 2$ and
$N_P= 2$. The integral is divergent and it yields a result of the form  
\be
A_{N^\pi}\sim \frac{1}{16\pi^2f_\pi^2}p^4 \times \frac{1}{\epsilon}.
\ee 
Dimensional regularization has been assumed.
The divergence can thus be absorbed by redefining the coefficient of the operators at 
${\cal O}(p^4)$ assuming that the regularization preserves chiral
invariance.

This counting works to all orders and IR divergences, that potentially could spoil it,
 are absent (Weinberg).
At each order in perturbation theory the divergences that arise can be 
eliminated by redefining the coefficients in the higher order operators, e.g.
\be 
\alpha_i\to \alpha_i + \frac{c_i}{\epsilon} 
\ee
Note that, in addition to the pure pole in $\epsilon$, 
logarithmic non-local terms necessarily appear. For instance in 
a two-point function they appear in the combination
\be
\frac{1}{\epsilon} + \log\frac{-p^2}{\mu^2}.
\ee
This comes about because pions are strictly massless in the chiral limit and 
thus a combination of momenta must necessarily normalize the $\mu^2$ that appears for dimensional
consistency in dimensional regularization.

The cut provided by the log is absolutely required by unitarity. Let us split
the scattering matrix $S$ in the usual way
\be 
S=I + iT.
\ee
The identity corresponds, obviously, to having no interaction at all. 

Unitarity implies
\be
S^\dagger S = I = I + i(T-T^\dagger) + T^\dagger T.
\nonumber\ee
\be
i(T-T^\dagger) = - T^\dagger T.
\ee
Thus $T$ must {\em necessarily} have an imaginary part. Pure powers of momenta
are real by construction. Thus the logs, that bring about a cut and an imaginary part,
are needed. Loops are essential, even for effective theories. There is no such thing as a
`classical effective theory' in a quantum theory.

To recapitulate,
the lowest-order, tree level contribution to pion-pion
scattering is $\sim \frac{p^2}{f_\pi^2}$.
The one-loop chiral corrections iare $\sim \frac{p^4}{16\pi^2 f_\pi^4}$. 
Thus the counting 
parameter in the loop (chiral) expansion is clearly
\be
\frac{p^2}{16\pi^2 f_\pi^2}.
\ee
Each chiral loop gives an additional power of $p^2$.

The counting can actually be extended to include small depatures from
the chiral limit, i.e. allowing for non-zero quark (hence pion) masses. If
${\cal O}(p^{2n})$ counts as $p^{2n}$,
soft breaking terms such as  
\be
\mu m {\rm Tr\;}  (U+U^\dagger)
\ee
give the pion a mass $m_\pi^2\sim m$. Therefore
$m$ counts as $p^2$ too. 

Note that all coefficients in the chiral lagrangian are nominally of ${\cal O}(N_c)$.
Loops are automatically suppressed by powers of $N_c$, because $f_\pi^2 \sim N_c$
appears in the denominator, but they are enhanced by
logs at low momenta as we just saw.

Chiral lagrangians are extremely successful. Their application to low-energy phenomenology
is nowadays standard and quite relevant. At any given order
in the derivative expansion a finite number of coefficients have to be
determined from experiment (or eventually lattice simulations), but then everything else
is known (with the precison given by the order retained in the derivative expansion). 
Even without knowing these coefficients one can find combinations of observables where the 
unknown coefficients drop. As an illustration we show recent fits to lattice 
data\cite{JN} using chiral lagrangians showing excellent agreement between their predictions and the 
numerical results; the point of course being that one can then  use the chiral lagrangian to 
extrapolate to a mass/energy regime unattainable by current numerical simulations. 

\begin{figure}[h]
\centering
\epsfysize=10cm
 \epsfbox{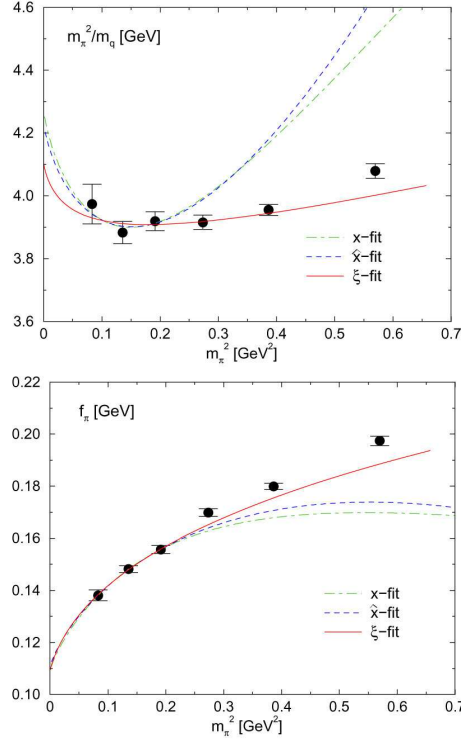}
 \caption{Recent fits to lattice data for light masses 
using chiral perturbation theory at the NLO. Extracted from 
reference \cite{JN}}  
\end{figure}

\section{The gravity analogy}

\noindent
The Einstein-Hilbert action shares several aspects with
the pion chiral lagrangian. Like the effective chiral lagrangian it is also a 
non-renormalizable theory (more on this latter). It is also described, considering 
the most relevant operator, by a dimension two operator containing in both cases
two derivatives of the dynamical variable. Both lagrangians contain necessarily
a dimensionful constant in four dimensions; the counterpart of  $f_\pi$ in the pion lagrangian
is the Planck mass $M_P$. Both theories are non-linear and, finally, both describe the
interactions of massless quanta. The Einstein-Hilbert action is
\be
{\cal L}= M_P^2 \sqrt{-g} {\mathcal R} + {\cal L}_{matter},
\ee
where
\be
\kappa^2 \equiv \frac{2}{M_P^2}= 32 \pi G
\ee
Indeed a cursory comparison with the expressions in the previous section shows that
$M_{P}$ plays a role very similar to $f_\pi$

As just mentioned
${\mathcal R}$ contains two derivatives of the dynamical
variable which is the metric $g_{\mu\nu}$
\be
{\mathcal R}_{\mu\nu} = 
\partial_\nu \Gamma^{\alpha}_{\mu \alpha} 
- 
\partial_{\alpha} \Gamma^{\alpha}_{\mu \nu} 
+ 
\Gamma^{\alpha}_{\beta \nu} \Gamma^{\beta}_{\mu\alpha}
- 
\Gamma^{\alpha}_{\beta \alpha} \Gamma^{\beta}_{\mu \nu}
\ee
\be
\Gamma^{ \gamma }_{ \alpha \beta } = \frac{1}{2} g^{\gamma \rho} \left( \partial_{\beta }
g_{\rho \alpha } + \partial_{ \alpha } g_{\rho \beta } - \partial_{\rho} g_{ \alpha \beta } \right)
\ee
\be
{\mathcal R}\sim \partial \partial g
\ee
In the chiral language, the Einstein-Hilbert action would be 
${\cal O}(p^2)$ i.e. most relevant, if we omit for a second the presence of the
cosmological constant which accompannies the identity operator. 

Arguably, locality,  symmetry and 
relevance in the RG sense (and not renormalizability) 
are the ones that single out Einstein-Hilbert action
in front of  e.g. ${\mathcal R}^2$.

Unlike the chiral lagrangian, the Einstein-Hilbert lagrangian (or extensions
thereof including higher derivatives) has a local gauge symmetry. Indeed,
gravity can be (somewhat loosely) described as the result of promoting a global symmetry (Lorentz) 
\begin{eqnarray}
x'^{a} & = & \Lambda^{a}_{~b} x^{b} \\
\eta_{a b} & = & \Lambda^{c}_{~a} \Lambda^{d}_{~b}
\eta_{c d} 
\end{eqnarray}
to a local one
\begin{eqnarray}
x^{\prime \mu}  =  x^{\prime \mu} (x) &\to &
dx^{\prime \mu}  =  \Lambda^{\mu}_{~\nu} (x) dx^{\nu}
\\
\bar{\Lambda}_{\mu} ^{~\nu} (x) & \equiv & \left[ \Lambda^{\mu}
_{~\nu}
(x) \right]^{-1} \\
\Lambda^{\mu} _{~\nu} \bar{\Lambda}_{\rho} ^{~\nu} & = &
\delta^{\mu}
_{\rho}
\end{eqnarray}
This can be acomplished if the basic field, the metric, is allowed to be a 
coordinate dependent field
transforming as
\begin{eqnarray}
g^{\prime}_{\mu \nu} (x^{\prime}) & = & \bar{\Lambda}_{\mu}
^{~\alpha}
\bar{\Lambda}_{\nu} ^{~\beta} g_{\alpha \beta} (x)  \\
d \tau^2 & = & g^{\prime}_{\mu \nu} (x^{\prime}) dx^{\prime \mu}
dx^{\prime \nu} = g_{\alpha \beta} (x) dx^{\alpha} dx^{\beta}
\end{eqnarray}
Fields transform as scalars, vectors, etc., under this change
\begin{eqnarray}
\phi^{\prime} (x^{\prime}) & = & \phi(x) \nonumber \\
A^{\prime \mu} (x^{\prime}) & = & \Lambda^{\mu}_{~\nu}(x) A^{\nu} (x)
\end{eqnarray}
This means that the gauge symmetry that is present in gravity,
unlike in the chiral lagrangian, will in practice reduce the number of degrees of freedom
that are observable at low energies for two reasons. One of the reasons of course is the
very existence of the gauge symmetry itself. For instance, describing a 
spin one particle (such as a massive photon)
with a four-vector is redundant; one of the four degrees of freedom 
completely decouples.

The other reason is easily understood just remembering 
what happens in the Standard Model of electroweak interactions where the global
symmetry is spontaneously broken down to $U(1)_{em}$, but because
of the $SU(2)_L \times U(1)_{Y}$ gauge invariance originally
present, all Goldstone
bosons disappear yielding, in turn, some massive modes that were previously
massless. The natural value for such masses is the Fermi scale ($\sim$ 250 GeV),
but in gravity it would undoubtedly be the Planck mass, disappearing in 
practice from the low energy dynamics. 

Einstein-Hilbert action has thus all the ingredients for being
an effective theory describing the long distance properties of some
unknown dyamics.

It is also natural to go one step further and ask whether
gravitons are just Goldstone bosons of some broken 
symmetry. We will have more to say about this possibility in the coming sections.

\subsection{Quantizing gravity}
Quantum corrections in gravity are 
analogous to the weak field expansion in pion physics
\be
U= I + i \frac{\tilde\pi}{f_\pi}+ . . . 
\ee
One writes
\begin{eqnarray}
g_{\mu \nu} & \equiv & \eta_{\mu \nu} + \kappa h_{\mu \nu} \,
 \\
g^{\mu \nu} &=& \eta^{\mu \nu} - \kappa h^{\mu \nu} + \kappa^2 h^{\mu
\lambda} h_{\lambda}^{~\nu} + \ldots
\end{eqnarray}
so in fact $\kappa$ plays the same role as $f_\pi^{-1}$.

The curvatures can likewise be expanded around a given background, say
$g_{\mu\nu} = \eta_{\mu\nu}$,
\begin{eqnarray}
{\mathcal R}_{\mu \nu} &=& {\frac{\kappa}{2}} \left[ \partial_{\mu} \partial_{\nu}
h^{\lambda} _{~\lambda} + \partial_{\lambda} \partial^{\lambda} h_{\mu
\nu} - \partial_{\mu} \partial_{\lambda} h^{\lambda} _{~\nu} -
\partial_{\lambda} \partial_{\nu} h^{\lambda} _{~\mu} \right] + {\cal O}
(h^2)  \\
{\mathcal R} &=& \kappa \left[ \Box h^{\lambda} _{~\lambda} - \partial_{\mu}
\partial_{\nu} h^{\mu \nu} \right] + {\cal O} (h^2).
\end{eqnarray}
Indices are raised and lowered with $\eta_{\mu \nu}$.
This can be done around any fixed 
background space time metric.

Green's functions do not exist without a gauge choice and it is
most convenient to use the so-called harmonic gauge where the Green functions
obey Poisson-like equations
\begin{equation}
\partial^{\lambda} h_{\mu \lambda} = \frac12 \partial_{\mu}
h^{\lambda} _{~\lambda}
\end{equation}
The well-known field equations
\be
{\mathcal R}_{\mu \nu} - \frac{1}{2} g_{\mu \nu} {\mathcal R} = - 8 \pi G T_{\mu \nu},
\qquad
\sqrt{g} T^{\mu \nu} \equiv - 2 {\frac{\delta}{ \delta g_{\mu \nu}} }
\left( \sqrt{g} {\cal L}_m \right)
\ee
reduce in this gauge to
\begin{equation}
\Box h_{\mu \nu} = -16 \pi G \left( T_{\mu \nu} - \frac{1}{2} \eta_{\mu
\nu} T^{\lambda} _{~\lambda} \right)
\end{equation}

The momentum space propagator is relatively simple in this gauge. Around
Minkowski space-time we obtain
\begin{equation}
i D_{\mu \nu \alpha \beta}  =  \frac{i}{q^2 + i \epsilon} P_{\mu \nu,
\alpha \beta} \qquad
P_{\mu \nu, \alpha \beta}  \equiv  \frac{1}{2} \left[ \eta_{\mu \alpha}
\eta_{\nu \beta} + \eta_{\mu \beta} \eta_{\nu \alpha} - \eta_{\mu \nu}
\eta_{\alpha \beta} \right]
\end{equation}
In addition one needs to include the gauge-fixing and ghost part.
Around an arbitrary backgroung $\bar g_{\mu\nu}$
\begin{equation}
{\cal L}_{gf} = \sqrt{ \bar{g}} \left\{ \left( D^{\nu} h_{\mu \nu} - 
\frac12 D_{\mu} h^{\lambda} _{~\lambda} \right) \left( D_{\sigma} h^{\mu
\sigma} - \frac12 D^{\mu} h^{\sigma} _{~\sigma} \right) \right\},
\end{equation}
\begin{equation}
{\cal L}_{gh} = \sqrt{ \bar{g}} \eta^{* \mu} \left[ D_{\lambda} D^{\lambda}
\bar{g}_{\mu \nu} - {\mathcal R}_{\mu \nu} \right] \eta^{\nu}
\end{equation}
It is plain that perturbative calculations in quantum gravity are 
quite difficult due to the proliferation of indices.

\subsection{Counterterms}

The following two results are well known and often quoted. The first one is
due to 't Hooft and Veltman, who computed the divergences in
pure gravity at the one loop level\cite{tv}. Without making use of the
equations of motion, the counterterms found by 't Hooft and Veltman in 
the harmonic gauge are
\begin{equation}
{\cal L}^{(div)}_{1 loop} = - \frac{1}{16 \pi^2 \epsilon} \left\{ 
\frac{1}{ 120}
{\mathcal R}^2 + \frac{7}{20} {\mathcal R}_{\mu \nu} {\mathcal R}^{\mu \nu} \right\}
\label{thooft}
\end{equation}

The second one is due to Goroff and Sagnotti\cite{gs} who performed 
a similar calculation at two loops. After using the equations of motion
\begin{equation}
{\cal L}^{(div)}_{2 loop} = - \frac{209 \kappa^2}{5760 (16 \pi^2 )}
\frac{1}{\epsilon} {\mathcal R}^{\alpha \beta}_{~~\gamma \delta} {\mathcal R}^{\gamma
\delta}_{~~\eta \sigma} {\mathcal R}^{\eta \sigma}_{~~\alpha \beta}
\end{equation}

It is less well appreciated that the two results are on a different footing.
The result of 't Hooft and Veltman is gauge dependent (it was computed in a particular
gauge --the harmonic gauge-- and it does not correspond to any physical observable, 
in particular the equations of motion have not been used). The counterterm
actually vanishes when the field equations in empty space are used ${\mathcal R}_{\mu\nu}=0$. 
The counterterm does give  a net divergence when $T_{\mu\nu} \neq 0$ and, therefore 
${\mathcal R}_{\mu\nu}\neq 0$, but the result is in principle 
incomplete as we will see below\cite{offshell}.

The one-loop counterterms computed by 't Hooft and Veltman, 
although historically quite relevant, are
thus largely irrelevant from the point of view of effective lagrangians
because they vanish on shell.

In de Sitter space, described by the action 
\begin{equation}
S=\frac{1}{16\pi G} \int dx \sqrt{-g} ({\mathcal R} - 2\Lambda)
\end{equation}
the counterterm structure was computed
by Christensen and Duff \cite{dc} in the 80's. A more detailed analysis
was performed later in \cite{kal,kkps}, where the gauge dependence of the
counterterms was clearly exposed
\begin{equation}
\Gamma^{(div)}_{eff}=
-\frac{1}{16\pi^2\epsilon }\int dx \sqrt{-g}
[c_1{\mathcal R}_{\mu\nu} {\mathcal R}^{\mu\nu}
+c_2\Lambda^2+ c_3 {\mathcal R}\Lambda + c_4{\mathcal R}^2].\label{gaugedep}
\end{equation}
The constants $c_i$ are actually gauge dependent and only
a combination of them is gauge invariant.

If we are interested in observables, the on-shell condition is to be imposed
on the counterterms of the effective theory (as in a derivative expansion
they will appear only at tree-level, see e.g. \cite{gl} for a discussion
on this).

Using the equations of motion (in absence of matter)
 ${\mathcal R}_{\mu\nu}= g_{\mu\nu}\Lambda$, the previous equation
reduces to the (gauge-invariant) on-shell expression \cite{kkps}
\begin{equation}
\Gamma^{(div)}_{eff}=
\frac{1}{16\pi^2\epsilon}\int dx \sqrt{-g}
\frac{29}{5}\Lambda^2.
\end{equation}
On the contrary,
if we set $\Lambda=0$ above, in (\ref{gaugedep}), and particularize to the
harmonic gauge, we reproduce the well-known 't Hooft
and Veltman divergence (\ref{thooft}).

Let us recapitulate.
Exactly as the chiral lagrangian, the  Einstein-Hilbert action requires an 
infinite number of counterterms
\be
{\cal L}= 
M_P^2 \sqrt{-g} {\mathcal R} +
\alpha_1 \sqrt{-g} {\mathcal R}^2 +
\alpha_2 \sqrt{-g} ({\mathcal R_{\mu\nu}})^2 + 
\alpha_3 \sqrt{-g} ({\mathcal R_{\mu\nu\alpha\beta}})^2 + \ldots
\ee
The divergences can be absorbed by redefining the coefficients just as 
done in the previous section for the pion effective lagrangian
\be
\alpha_i \to \alpha_i + \frac{c_i}{\epsilon}
\ee
Power counting in gravity appears, at least superficially, quite similar to
the one that can be implemented in pion physics.
Of course, the natural expansion parameter is a tiny number in normal circumstances,
namely
\be
p^2/16\pi^2 M_{P}^2\quad {\rm or}\quad 
\nabla^2/16\pi^2 M_{P}^2,\qquad {\mathcal R}/16\pi^2 M_{P}^2
\ee
Because of this, Donoghue has termed the ffective action of gravity the most 
effective of all effective actions!

\section{Why we need genuine loop effects and power counting}

Consider the following {\em generic} ${\mathcal R}^2$ correction to
the Einstein-Hilbert action 
\begin{equation}
{\cal L} = \frac{2}{\kappa^2} {\mathcal R} + c{\mathcal R}^2+ {\cal L}_{matter}.
\end{equation}
The corresponding equation of motion for a perturbation around 
Minkowski is (recall that we write $g=\eta+h$) 
\begin{equation}
\Box h + \kappa^2 c^2 \Box \Box h = 8 \pi G\, T. 
\end{equation}
The Green function for this equation has the form
\begin{eqnarray}
G(x) & = & \int \frac{d^4q}{ (2 \pi)^4 } \frac{e^{-iq \cdot x}}{q^2
+ \kappa^2 c
q^4} \\
& = & \int \frac{d^4 q}{(2 \pi )^4} \left[ \frac{1}{ q^2} - \frac{1}{q^2 +
{1/\kappa^2 c}} \right] e^{-iq \cdot x}
\end{eqnarray}

Taken at face these higher order terms would lead 
to a correction to Newton's law
\begin{equation}
V(r) = - G m_1 m_2 \left[ \frac{1}{ r} - 
\frac{e^{-r/ \sqrt{\kappa^2 c}}}{r}\right]
\end{equation}
Experimental bounds indicate $c < 10^{74}$; that is, no bound at all
in practice. This is of course a consequence of the 'effectiveness' of the effective action
of gravity.  If $c$ was a
reasonable number there would be no effect on any observable physics
at terrestrial scales.
Note that if $c \sim 1, \sqrt{\kappa^2 c} \sim 10^{-35}m$.  
The curvature is so small that ${\mathcal R}^2$ terms are 
completely irrelevant at ordinary scales.

However using the full solution of the wave equation is {\it not} compatible
with the effective lagrangian philosophy and the power counting
it embodies because higher orders in $\kappa$ are 
sensitive to higher curvatures we have not considered.

The leading behaviour of the correction is
\begin{equation}
\frac{e^{-r/ \sqrt{\kappa^2c}}}{r} \rightarrow 4\pi
\kappa^2c \delta^3 (\vec{r}).
\end{equation}
In momentum space this translates into
\begin{equation}
\frac{1}{q^2 + \kappa^2c q^4} = \frac{1}{q^2} - \kappa^2c + \cdots
\end{equation}
Thus the 'correction' to Newton's law coming from the ${\mathcal R}^2$ correction is
\begin{equation}
V(r) = -Gm_1 M_2 \left[\frac{1}{r} + 128 \pi^2 G c \delta^3
(\vec{x}) \right],
\end{equation}
which is totally unobservable, even as a matter of principle.

Of course, apart from the divergences, there are finite pieces (not universal, due to
the renormalization ambiguities, choice of different substraction methods, etc. ) and, most
importantly, {\it non-local} pieces.  Indeed in dimensional regularization
 we get at the one-loop level
\be
\frac{1}{\epsilon} + \log\frac{-p^2}{\mu^2}
\ee
Or, in position space,
\be
\frac{1}{\epsilon} + \log\frac{\nabla^2}{\mu^2},
\ee
where
$\nabla$ has to be the covariant derivative on symmetry grounds, $\nabla^2$ 
reducing to $-p^2$
in flat space-time. These non-localities  are due to the propagation of strictly
massless non-conformal modes, such as the graviton itself. Therefore 
they are unavoidable
in quantum gravity. Notice that the coefficient is predictable; it depends entirely
on the infrared properties of gravity.

\section{Quantum corrections to Newton law}

Let us use the 'chiral counting' arguments to derive the relevant quantum 
corrections to Newton's law (up to a constant). The propagator at tree level,
that we symbolically write as
\be
\qquad\frac{1}{p^2}\, ,
\ee
gets modified by the  one-loop 'chiral-like' corrections to 
\be
\qquad \frac{1}{p^2}(1 + A\frac{p^2}{M_P^2} + B \frac{p^2}{M_P^2} \log p^2).
\ee
Of course the last expression is also symbolic.

Consider now the interaction of a point-like particle with an static source ($p^0=0$) 
and let us Fourier transform the previous expression for the loop-corrected 
propagator in order to get the potential
in the non-relativistic limit. We use
\be
\int d^3 x \exp (i\vec p \vec x) \,\frac{1}{p^2} \sim \frac{1}{r}
\qquad
\int d^3 x \exp (i\vec p \vec x) \,1  \sim \delta({\vec x})
\ee
\be
\int d^3 x \exp (i\vec p \vec x) \, \log{p^2} \sim \frac{1}{r^3}
\ee
Thus the quantum corrections ro Newton's law are of the form
\be
\frac{GMm}{r}( 1 + C \frac{G\hbar}{r^2}+ \ldots ).
\ee
We have restored for a moment $\hbar$. Let us check dimensions.
We note that  
\be
\left[\frac{Gm}{c^2} \right]= L,   \qquad \left[\frac{G\hbar}{c^3}  
\right]= L^2
 \ee
\noindent
so $C$ is a pure number. In addition there are post-newtonian (but classical) 
corrections that are not discussed here.

A long controversy regarding the value of $C$ exist in the literature.
Donoghue, Muzinich, Vokos, Hamber, Liu, Bellucci, Khriplovich, Kirilin,
Holstein, Bjerrum-Bohr and others have contributed\cite{khri,donbohr,qcorrs} to the determination
of $C$. 
The result widely accepted as the correct one\cite{bohr} is obtained 
by considering the inclusion
of {\it quantum} matter fields (a scalar field actually)
and considering all type of loops

The relevant set of Feynman rules is
\begin{eqnarray}
\tau_{\mu \nu} & = & -\frac{i\kappa}{2} \left( p_{\mu}
p^{\prime}_{\nu} +
p^{\prime}_{\mu} p_{\nu} - g_{\mu \nu} [p \cdot p^{\prime} - m^2]
\right)
\\
\tau_{\eta \lambda, \rho \sigma} & = & \frac{i \kappa^2}{2} \left\{
I_{\eta \lambda, \alpha \delta} I^\delta _{~\beta, \rho \sigma} \left(
p^{\alpha} p^{\prime \beta} + p^{\prime \alpha} p^{\beta} \right) \right. 
\\
& & - \frac12 \left( \eta_{\eta \lambda} I_{\rho \sigma,
\alpha \beta} +  \eta_{\rho \sigma} I_{\eta \lambda, \alpha \beta} \right)
p^{\prime \alpha}  p^{\beta}  \nonumber \\
& & \left. - \frac12 \left( I_{\eta \lambda, \rho \sigma} - \frac12
\eta_{\eta \lambda} \eta_{\rho \sigma} \right) [p \cdot p^{\prime} - m^2 ]
\right\},
\end{eqnarray} 
with
\begin{equation}
I_{\mu \nu, \alpha \beta} \equiv \frac12 [ \eta_{\mu \alpha} \eta_{\nu
\beta} + \eta_{\mu \beta} \eta_{\nu \alpha} ]
\end{equation}
The first Feynman rule corresponds to a matter-matter-1-graviton vertex, while
the second one describes the matter-matter-2-graviton interaction. Actually 
the interaction with matter always takes place via the energy-momentum tensor. Note
that (quantum) matter does propagate inside loops. Please note that very heavy
(matter) degrees of freedom do not necessarily decouple from quantum 
corrections as the coupling itself to gravity depends on the mass. 

In addition one needs the 3-graviton interaction vertex which is described by
 quite a lengthy 
expression and shall not be given here. It can be found in \cite{don}.

Then, in a rather informal but otherwise obvious notation, the calculation
of the local counterterms gives \cite{khri}
\be
{\cal L}_{\mathcal RR}= \frac{1}{3849 \pi^3 r^3} 
(42 {\mathcal R}_{\mu\nu}{\mathcal R}^{\mu\nu}
+ {\mathcal R}^2)\ee
\be{\cal L}_{{\mathcal R}T}= -\frac{\kappa}{8 \pi^2 r^3} (3{\mathcal R}_{\mu\nu} T^{\mu\nu}
- 2 {\mathcal R}T)\ee
\be{\cal L}_{TT}= \frac{\kappa^2}{60 \pi r^3} T^2\ee

At this point one can make use of the lowest order equations of motion to simplify the
counterterm structure
\be
{\mathcal R}_{\mu \nu} - \frac12 g_{\mu \nu} {\mathcal R} = - 8 \pi G T_{\mu \nu}
\ee
\be
\Rightarrow \quad {\cal L}_{total}= - \frac{\kappa^2}{60 \pi r^3} (138
T_{\mu\nu} T^{\mu\nu} -31 T^2)
\ee
Particularizing now to the case of a point-like mass, we get the final
result for $C$, which is positive in sign: gravity is more atractive at long distances than
predicted by Newton's law (although the difference is of course extremely tiny)
\be
C= \frac{41}{10\pi}
\ee
What happens for {\it classical} matter, e.g. a cloud of dust, is
in our view still an open problem.

There are in the literature definitions of an ``effective'' or ``running'' Newton constant \cite{jp,dl}. 
A class of diagrams is identified that dresses up $G$ and turns it into 
a distance (or energy)-dependent constant $G(r)$. Unfortunately it is not clear
to us that these definitions are gauge invariant; only physical observables 
(such as a scattering matrix) are guaranteed to be. So caution should be adopted here, 
although 
the renormalization-group analysis derived from this ``running'' coupling constant are of course 
very interesting.

\subsection{Power counting in gravity}

Let us try to establish a counting analogous to the one we did for the
pion chiral lagrangian. Some of the counting rules are obvious, others require
a little thought. Let us indicate them, again symbolically

\smallskip
$\bullet$  3-graviton coupling:  $\sim \kappa  q^2$

\smallskip
$\bullet$  4-graviton coupling:  $\sim \kappa^2  q^2$

\smallskip
$\bullet$  (On-shell) matter-- 1-graviton coupling:  $\sim \kappa m^2$

\smallskip
$\bullet$  (On-shell) matter-- 2-graviton coupling:  $\sim \kappa^2 m^2$  

\smallskip
$\bullet$ Graviton propagator: $\sim \frac{1}{q^2}$

\smallskip
$\bullet$ Matter propagator $\sim \frac{1}{q^2-m^2}$

\smallskip
If we iterate, for example, the 4-graviton vertex to produce a one loop diagram we 
shall obtain ($p_i$ are external momenta and $q=p_1+p_2$)
\be{{\cal M}_{loop}\sim\kappa^4\int\frac{d^4l}{(2\pi)^4}
\frac{(l-p_1)^2(l-p_2^2)^2}{l^2(l-q)^2}}
\ee
If this
loop integral is regularized dimensionally, which does not introduce
powers
of any new scale, the integral will be represented in terms of the
exchanged momentum to the appropriate power. Thus we have
\be
{{\cal M}_{loop}\sim\kappa^4\, q^4}
\ee

When matter 
fields are included in loops the situation is more subtle, in particular
for large masses in the non-relativistic limit. Let us see why.
If we compute the
tree level result for matter-matter scattering the result is
\be
{\cal M}_{tree}=\kappa^2\frac{m_1^2m_2^2}{q^2}
\ee
Note that this is not yet the potential, hence the unfamiliar power of the masses
in the numerator. Iterating this expression to form a loop
one encounters internal lines where a matter field propagates. This propagator
has a denominator of the form $(k-q)^2-m^2$ that on shell and for large masses
in the non-relativistic limit will behave as $mq$. Therefore one gets
\be{\cal M}_{loop}\sim\kappa^4m_1^4m_2^4\int{d^4l}
\frac{1}{m_1(l+p)}\times\frac{1}{m_2(l+p^\prime)}\times
\frac{1}{(l+q^\prime)^2}\times\frac{1}{(l+q)^2}
\ee
which by the same reasoning as before is
\be
{\cal M}_{loop}\sim\kappa^4\frac{m_1^3m_2^3}{q^2}\sim
\kappa^2\frac{m_1^2m_2^2}{q^2}\times\kappa^2m_1m_2
\ee
Here the expansion parameter appears to be $\kappa^2m^2$ that does not
seem compatible with the
`chiral' expansion arguments.
 
This issue has been studied by some detail by Donoghue and Torma \cite{dt} who concluded
that
\be
{\cal M}_{(N^m_E,N^g_E)} \sim q^D
\ee
where
\be
 D=2-\frac{N^m_E}{2}+2N_L-N^m_V+\sum_n{(n-2)N^g_V[n]}+
\sum_l{l\cdot N^m_V[l]},
\ee
being $N_E$, $N_L$ and $N_V$ the number of external fields, loops and
vertices, respectively, and the superindex refering to whether they are matter 
or gravity fields.
If we disregard matter vertices this is identical 
to Weinberg's result for chiral theories \cite{wein}, who concluded
that the power counting expansion is sound for the pion 
effective lagrangian.

However the negative $N^m_V$ term appearing in $D$ is potentially
dangerous. Although no general proof exists yet, Donoghue has been able to prove
cancellation of the dangerous terms at the one-loop level 
except for the terms leading to $1/r$ corrections (classical, non-linear).
The issue is, to our knowledge, still not fully solved.

We conclude with a final comment concerning the use of the
equations of motion.
In chiral lagrangians they allow us to get rid of redundant 
operators. For instance, taking into account that from the
lowest order lagrangian results the following Euler-Lagrange
equation
\be
U\Box U^\dagger -(\Box U) U^\dagger =0
\ee
we can set, at the next order in the chiral expansion,
\be
{\rm Tr\;} U\Box U^\dagger \to 0
\ee

However, note
that in gravity, the equation of motion mixes
terms of different `chiral' order
\be
{\mathcal R}_{\mu \nu} - \frac12 g_{\mu \nu} {\mathcal R} = - 8 \pi G T_{\mu \nu}-
 g_{\mu\nu}\Lambda
\ee

For instance, it is incorrect to use 
\be
{\mathcal R}_{\mu \nu} =
 g_{\mu\nu}\Lambda
\ee
in 't Hooft and Veltman calculation, even
if $\Lambda$ is generated by the v.e.v. of some scalar field (as long as is spatially constant
and does not vary with time) which is induced by some (dimension four) matter sector. 
It just does not
reproduce the de Sitter result.

\section{Cosmological implications}

The quantum corrections to Newton's law emerge from the universal
non-local corrections to the effective action. They constitute a 
direct test of the quantum nature of gravitation, putting this theory
on an equal footing to other quantum field theories. They are thus conceptually 
extremely important, but it is hard to imagine how one could measure such 
a tiny effect. Can these non-local quantum corrections be relevant, or at least observable,
in a cosmological setting?

We are concerned here about universal non-local quantum corrections to
the Einstein-Hilbert lagrangian that take the form (again symbolically)
 \be
\frac{1}{16\pi^2M_{P}^2}{\mathcal R}[\log \nabla^2] {\mathcal R}.\label{nlocal}
\ee
There are two reasons why  such apparently hopelessly small
corrections might 
be relevant in a cosmological setting

--- Curvature was much larger at early stages of the universe: in 
a de Sitter universe ${\mathcal R}\sim H^2$, $ H^2= 8\pi G V_0/3$, 
$H\le 10^{13}$ GeV 
(present value is $10^{-42}$ GeV). 

--- Logarithmic
non local term corresponds to an interaction between geometries 
that is long-range in time, an effect that does not have an 
easy classical interpretation.

Please note that the above non-local contributions  are 
totally unrelated to the so-called $f({\mathcal R})$ models. They
are present and unambigously calculable in the quantum theory.
It should be mentioned here too that
somewhat related  non-localities (but at the two loop level)
were studied by Tsamis and Woodard long ago\cite{tw}. They turn out to 
slow down the rate of
inflation.

For the purpose of the present discussion let us spell out our
conventions
\begin{equation}
S=\frac{1}{16\pi G} \int dx \sqrt{-g} ({\mathcal R} - 2\Lambda)
+ S_{matter},\quad
 {\mathcal R}_{\mu\nu} - \frac{1}{2}{\mathcal R} g_{\mu\nu}= -8\pi G T_{\mu\nu}  
- \Lambda g_{\mu\nu}
\end{equation}

Quantum corrections to the Einstein-Hilbert action were originally computed by
't Hooft and Veltman in the case of vanishing 
cosmological constant \cite{tv},
and by Chistensen and Duff for a de Sitter background\cite{dc}.
The key ingredient we shall need is the divergent part of the one-loop
effective action. Setting
$d=4 + 2\epsilon$  
\begin{equation}
\Gamma^{div}_{eff}=
-\frac{1}{16\pi^2\epsilon }\int dx \sqrt{-g}
[c_1{\mathcal R}_{\mu\nu} {\mathcal R}^{\mu\nu}
+c_2\Lambda^2+ c_3 {\mathcal R}\Lambda + c_4{\mathcal R}^2].
\end{equation}
The constants $c_i$ are actually gauge dependent as has already been mentiones and only
a combination of them is gauge invariant. This is clearly
discussed in \cite{kal,kkps}.

Using the equations of motion (in absence of matter)
 ${\mathcal R}_{\mu\nu}= g_{\mu\nu}\Lambda$, the previous equation
reduces to the (gauge-invariant) on-shell expression
\begin{equation}
\Gamma^{div}_{eff}=
\frac{1}{16\pi^2\epsilon}\int dx \sqrt{-g}
\,\frac{29}{5}\Lambda^2.
\end{equation}

If we set $\Lambda=0$ above, we get the well-known 't Hooft
and Veltman divergence, that in the so-called minimal gauge is
\begin{equation}
\Gamma^{div}_{eff}=
-\frac{1}{16\pi^2\epsilon}\int dx \sqrt{-g}
\,[\frac{7}{20}{\mathcal R}_{\mu\nu} {\mathcal R}^{\mu\nu}
+ \frac{1}{120}{\mathcal R}^2].
\end{equation}
If the equations of motion are used in the absence of matter this divergence
is absent. 

Let us now try to investigate to what extent the non-local quantum 
corrections to the effective action, represented by (\ref{nlocal}) can
modify the evolution of the cosmological scale factor in
a Friedman-Robertson-Walker universe.

In what follows we summarize the results presented in \cite{ce,emv}.
For the sake of discussion, we shall begin by considering here a simplified 
effective action that includes 
only terms containing the scalar curvature 
\begin{align}
 \notag
 S &= \kappa^2 \left( \int dx \sqrt{-g} \mathcal{R} 
+ \tilde{\alpha} \int dx \sqrt{-g} \mathcal{R} \ln ( \nabla^2 / \mu^2 ) 
\mathcal{R} + \tilde{\beta} \int dx \sqrt{-g} \mathcal{R}^2 \right)
 \\
 &\equiv
 \kappa^2 \left( S_1 + \tilde\alpha S_2 + \tilde\beta S_3 \right),
 \label{effaction}
\end{align} 
where $\kappa^2=M^2_{P}/16\pi=1/16\pi G$ and $\mu$ is the subtraction
scale.
The coupling $\tilde\beta$ is $\mu$ dependent in such a way that 
the total action $S$ is $\mu$-independent. 

Note that

--- The value
of $\tilde\beta$ is actually dependent on the UV structure of the
theory (it contains information on all the modes -massive or not- that
have been integrated out)

--- The value of $\tilde\alpha$ is unambiguous: 
it depends only on the IR structure of gravity (described
by the Einstein-Hilbert Lagrangian)
and the massless (nonconformal) modes. 

In conformal time
\begin{equation}
g_{\mu\nu}=a^2(\tau)\eta_{\mu\nu},\ 
\mathcal{R}=6 \frac{{a''(\tau)}}{a^3(\tau)},\
\sqrt{-g}=a^4(\tau).
\end{equation}

We first obtain the variation of the local part
\begin{equation}
 \frac{\delta S_1}{\delta a(\tau)} = 12 a'',
\quad
 \frac{\delta S_3}{\delta a(\tau)} = 
72 \left( -3 \frac{(a'')^2}{a^3} - 4 \frac{a' a'''}{a^3} 
+
 6 \frac{(a')^2 a''}{a^4} + \frac{a^{(4)}}{a^2} \right).
\end{equation}

In order to obtain the variation of the non-local (logarithmic piece) we need to
compute 
\be
\langle x\vert \log \nabla^2 \vert y\rangle,
\ee
where in conformal coordinates
\begin{equation}
\nabla^2={a}^{-3}\Box\, a +\frac{1}{6} \mathcal{R}.
\end{equation}
To the order we are computing we can
neglect the $\mathcal{R}$ term in the previous equation and 
commute the scale factor $a$ with the flat d'Alembertian 
\begin{equation}
\nabla^2=\left(\frac{a}{a_0}\right)^{-2}\Box
\end{equation}
Where $a_0=a(0)$. With this rescaling (absorbable in $\tilde{\beta}$), 
at $\tau=0$
the d'Alembertian in conformal space matches with the Minkowskian one.

We can now separate $S_2$ in turn into a local and a genuinely non-local piece
\be
S_2=\int dx\sqrt{-g}\ \left(-2\mathcal{R}\ln(a)\mathcal{R}+\mathcal{R}\ln(\Box/\mu^2)\mathcal{R}\right)
\equiv S_2^I+S_2^{II}.
\ee
\begin{equation}
\begin{aligned}
\frac{\delta S_2^I}{\delta a(\tau)}  
=&
-72\left\lbrace 
\frac{(a')^2 a''}{a^4}  \left[ 12 \ln a - 10 \right]\right.\\
& + 
\frac{a' a'''}{a^3} \left[ -8 \ln a + 4 \right]
+
\frac{(a'')^2}{a^3} \left[ -6 \ln a + 2 \right]
+
\frac{a^{(4)}}{a^2} 2 \ln a
\rbrace
\end{aligned}
\end{equation}
Finally we have to compute 
\begin{equation}
\langle x\vert \ln \Box\vert y\rangle = \lim_{\epsilon\to 0}
\frac{1}{\epsilon} \langle x\vert \Box^\epsilon \vert y\rangle
-\frac{1}{\epsilon} \langle x\vert y\rangle
\end{equation}
The (covariant) delta function is in one-to-one correspondence with the counterterm.
The Green's function we are interested will be
\be
\sim \frac{1}{\vert x -y \vert^{4+2\epsilon}}.
\ee
After integration of $\vec x -\vec y$ we get
\be
\sim \frac{1}{\vert t - t^\prime\vert^{1+2\epsilon}}.
\ee
So
\begin{equation}
 S_2^{II} = 36 \int d\tau \frac{a''(\tau)}{a(\tau)} 
\int_0^{\tau} d\tau' \frac{1}{\tau-\tau'} \frac{a''(\tau')}{a(\tau')}.
\end{equation}
Note the limits of integration ensuring causality. Technically speaking we are using
here the in-in effective action and not the in-out one that would be appropriate for
a scattering process.

The variation of $S_2^{II}$ is
\begin{equation}
\begin{aligned}
 \frac{\delta S_2^{II}}{\delta a(\tau)} =&
36 \left\lbrace 
\left[ 2 a^{-3}(\tau)\left(a'(\tau)\right)^2 
- 2 a^{-2}(\tau) a''(\tau) \right] 
\int_0^\tau d\tau' \frac{1}{\tau - \tau'} \frac{a''(\tau')}{ a(\tau')}
\right.\\
& - 2 a^{-2}(\tau) a'(\tau) \frac{\partial}{\partial \tau} 
\left(\int_0^\tau d\tau' \frac{1}{\tau - \tau'} \frac{a''(\tau')}{ a(\tau')}
\right)\\ 
& + 
a^{-1}(\tau) \frac{\partial^2}{\partial \tau^2} 
\left( \int_0^\tau d\tau' \frac{1}{\tau - \tau'} \frac{a''(\tau')}{ a(\tau')} 
\right) 
\rbrace.
\end{aligned}
\end{equation}
In the spirit of effective Lagrangians we would
obtain first the lowest order equation of motion from $S_1$ and plug it in
$\tilde\alpha (S_2^I + S_2^{II}) + \tilde\beta S_3$.
As can be seen by inspection, quantum corrections
act as an external driving force superimposed to Einstein
equations. 

In a FRW universe without matter and with zero cosmological constant the
non-local pieces are actually zero (i.e. there are no log terms)
when one considers physical observables and the equations of motion are used. 
Therefore the toy model we have considered is not realistic, but it has served us to
develop our tools.

Let us now move to the more physically relevant case of
a de Sitter universe. 
The relevant one-loop corrected effective action is 
\be
S=\frac{1}{16\pi G} \int dx \sqrt{-g} ({\mathcal R}-2\Lambda) 
+\frac{1}{16\pi^2}\int dx \sqrt{-g} \frac{29}{5} 
\Lambda \ln\frac{\nabla^2}{\mu^2} 
\Lambda 
\ee
\be
+ {\rm local ~terms~of~} {\cal O}(p^4).\ee
We write $S$ as
\be
S \equiv  \kappa^2\left( \int dx \sqrt{-g} ({\mathcal R}-2\Lambda)
 + \tilde\alpha S_2\right),
\ee
with
\begin{equation}
 \tilde\alpha= \frac{G}{\pi}\times \frac{29}{5}.
\end{equation}
We split $S_2$ in two parts
\be
 S_2^{I} = -2 \int dx \sqrt{-g} \Lambda^2\ln(a), \qquad
 S_2^{II} = \int dx \sqrt{-g} \Lambda \ln(\square/\mu^2) \Lambda, 
\ee
and obtain the corresponding variations following the 
method outlined previously
\begin{equation}
 \frac{\delta S_2^I}{\delta a(\tau)} = -2 \Lambda^2 a^3(\tau) \left[4\ln(a(\tau)) + 1\right],
\end{equation}
\begin{equation}
 \frac{\delta S_2^{II}}{\delta a(\tau)} =  
2 \Lambda^2 a(\tau) \int_0^\tau d\tau' a^2(\tau') \frac{\mu^{-2\epsilon}}{\vert \tau - \tau' \vert^{1+2\epsilon}}.
\end{equation}
The equation of motion will be 
\begin{equation}
12 a^{\prime\prime}(\tau) - 8 \Lambda a^3(\tau) 
+\tilde\alpha 
\frac{\delta S_2}{\delta a(\tau)}=0
\end{equation} 
which at lowest order is just
\begin{equation}
12 a^{\prime\prime}(\tau) - 24 H^2 a^3(\tau)= 0,\qquad H^2=\Lambda/3
\end{equation}
The lowest order solution (with $a(0)=1$) is
\begin{equation}
a_I(\tau)=\frac{1}{ 1-H \tau}
\end{equation}
The final step  
is to plug the $0$-th order solution $a_I(\tau)$ into the 
variation of $S_2$ and recalculate the solution for $a(\tau)$. 
Note that we use a 
perturbative procedure
is of course only valid as long as 
the correction is small compared to the unperturbed
solutions.

We introduce a variable $s$ defined  $a_I(\tau) = e^s$.  Then
$s$ counts the number of e-folds 
\begin{equation}
\frac{\delta S_2^I}{\delta a(\tau)} = 
-2 \Lambda^2 e^{3s} \left[4s + 1\right]
\qquad
\frac{\delta S_2^{II}}{\delta a(\tau)}
  = 2 \Lambda^2 e^s I(s)
\end{equation}
and the equation of motion reads
\begin{equation}
 a''(s) + a'(s) - 2 e^{-2s}a^3(s) = 
\frac{3}{2} \tilde{\alpha} H^2 \left( - e^{s} (1 + 4s) + e^{-s} I(s) \right),
\end{equation}
where $I$ is 
\begin{equation}
 I(s) =  \ln\left(\frac{\mu}{H} (1 - e^{-s})\right) e^{2s} + e^s(1 - e^s - se^s),
\end{equation}
and the equation to solve is 
\begin{equation}
  a''(s) + a'(s) - 2 e^{-2s}a^3(s) = 
\frac{3}{2} \tilde{\alpha} H^2 \left[ - (5 s + 2) e^{ss} + 1 +
   e^{s} \ln\left(\frac{\mu}{H} (1 - e^{-s}) \right) \right]
\end{equation}
Note that $\tilde\alpha$ appears only in the combination
$\tilde\alpha H^2$. Since there are $H$ large uncertainties
in $H$ in practice only the sign of $\tilde\alpha$ is relevant. In addition,
there is some ambiguity associated to the choice of the renormalization
scale that appears in the combination $\ln(\mu/ H)$. This shown in Figure 2.

\begin{figure}[h]
\centering
\epsfysize=5cm
 \epsfbox{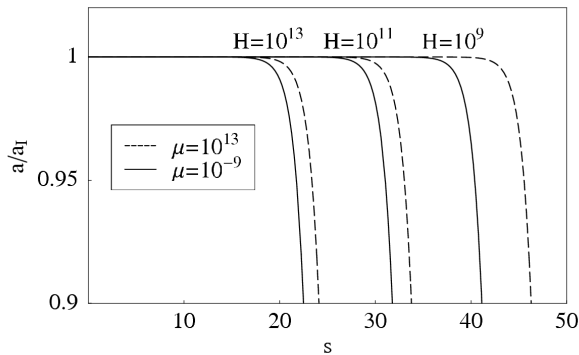}
 \caption{The scale factor relative to the inflationary expansion 
for different values of $\mu$ and  $H$ (all units are $\hbox{GeV}$).
We can see that 
the curves present a very similar behaviour for the different values shown, 
though a higher value of $H$ leads earlier to deviations from the usual 
inflationary expansion. Higher values of $\mu$ also have this effect,
 which is larger as $H$ increases. In fact, if we considered 
values of $\mu/H$ large enough (but not relevant physically), 
the logarithm term would become dominant and the deviation would be positive. }  
\end{figure}

Let us now assume that $a=a(\tau, \vec x)$; i.e. we allow for some space inhomogeneities. 
Then
\be
 \frac{\delta S_2^{II}}{\delta a(\tau,\vec x)} \sim \Lambda^2 a(\tau,\vec x) 
\int_0^\tau d\tau'd^3\vec y a^2(\tau',\vec y) \frac{\mu^{-2\epsilon}}{\vert 
x - y \vert^{4+2\epsilon}}.
\ee

This corresponds to new correlations of a quantum nature between different 
points. The consequences of this term have not been fully investigated yet.

\section{Gravity as a Goldstone phenomenon}

We have given in the previous sections arguments why the Einstein-Hilbert 
action could be viewed
as the most relevant term, in the sense of the renormalization-group, 
of an effective theory. 

Let us review them:

--- Dimensionful coupling constant ($M_{P}\sim f_\pi$)

--- Derivative couplings  ( $\sqrt{-g} {\mathcal R} \sim g\partial\partial g$)

--- Choice of action based on RG criteria of relevance, not on renormalizability (unlike
Yang-Mills)

--- Power counting anologous to ChPT

--- Massless quanta ($\pi \leftrightarrow g_{\mu\nu}$)

--- Existence of a global symmetry to be broken (see below)

Here we want to pursue this line of thought further.
As an entertainment, without making any particularly strong
claim of relevance, we shall investigate a formulation inspired as much as
possible in the chiral symmetry breaking of QCD. It has the following
characteristics:

--- No {\it a priori} metric, only affine connection is needed (parallelism)

--- Lagrangian is manifestly independent of the metric

--- Breaking is triggered by a fermion condensate

A different model along these lines was considered some time ago by Russo 
and coworkers\cite{ru}.

We seek inspiration in the effective lagrangians of QCD at long distances.
A successful model for QCD is the so-called chiral quark model. Consider
the matter part lagrangian of QCD with massless quarks (2 flavours)
\be
{\cal L}= i\bar \psi \not\! \partial \psi 
= i\bar \psi_L \not\! \partial \psi_L
+i\bar \psi_R \not\! \partial \psi_R.
\ee
This theory has a global $ SU(2)\times SU(2)$ symmetry that forbids
a mass term $M$.

However after chiral symmetry breaking pions appear and they must be
included in the effective theory. Then it is possible to add
the following term
\be
- M \bar\psi_L U \psi_R -M \bar\psi_R U^\dagger\psi_L, 
\ee 
that is invariant under the full global symmetry
\be \psi_L\to L\psi_L,\qquad \psi_R \to R\psi_R,\qquad
U\to LUR^\dagger.
\ee

Chiral symmetry breaking is also characterized by the presence of 
a fermion condensate 
\be
<\bar\psi\psi>\neq 0.
\ee
In order to determine whether the condensate is zero or not one is to solve
a `gap'-like equation in some modelization of QCD, or on the lattice.
The final step is to integrate out the fermions using
the self-generated effective mass as an infrared regulator. This reproduces the chiral 
effective lagrangian discussed in the
beginning of the lectures, although the low-energy constants $\alpha_i$ obtained 
in this way are not necessarily the
real ones, as the chiral quark model is only a simplification of QCD and not the real thing.

There is only one possible term bilinear in fermions 
that is invariant under Lorentz $\times$ {\it Diff} 
\be 
\bar{\psi}_a \gamma^a \nabla_\mu \psi^{\mu}
\ee
To define $\nabla$ we only need an affine connection 
\be
\nabla_{\mu} \psi^{\mu} = \partial_\mu \psi^{\mu} + i  \omega^{a b}_{\mu}
  \sigma_{a b} \psi^{\mu} + \Gamma^{\nu}_{\mu \nu} \psi^{\mu}
\ee
Note that no metric is needed at all
to define the action if we assume that $\psi^\mu$ behaves
as a contravariant spinorial vector density under {\it Diff}. Then,
$\Gamma^\mu_{\nu\rho}$ does not enter, only the spin connection.
If we keep this spin connection fixed, i.e. we do not consider it
to be a dynamical field for the time being, there is no invariance under 
general coordinate transformations, but only under the global group
$SO(d)\times GL(d)$ (assuming an Euclidean signature)\footnote{We recommend 
the reader to follow the discussion presented by Percacci\cite{per} in these
same proceedings}.

Eventually
we would like to find a non trivial condensate such as
\be
<\bar \psi_a \psi^\mu> \sim e_a^\mu.
\ee
In the absence of the (so far) external connection, we expect
a constant value for $e_a^\mu$ (note that the constant of proportionality
has dimensions of mass if we take $e_a^\mu$ to be dimensionless). 
It is of course irrelevant in which
direction it points; all the vacua will be equivalent. If the
condensate appears one can always choose $e_a^\mu = \delta_a^\mu$ without loss
of generality. We shall interpret $e_a^\mu$ as the (inverse) n-bein.
Note that once a dynamical value for $e_a^\mu$ is generated we can write terms
such as $M \bar \psi_a e^a_\mu \psi^\mu$, where $e^a_\mu$ (the n-bein) is 
defined by $e^a_\mu e_b^\mu = \delta^a_b$. Of course one can introduce
quantities such as $g^{\mu\nu}= e_a^\mu e_b^\nu \delta_{ab}$ and its
inverse $g_{\mu\nu}$ defined by $g_{\mu\nu}g^{\nu\rho}=\delta_\mu^\rho$.

Note that a large number of Goldstone bosons are produced. The original symmetry group
$G=SO(d)\times GL(d)$ has $\frac{d(d-1)}{2}+d^2$ generators. After the breaking 
$G\to H$, with $H=SO(d)$, which has a total of $\frac{d(d-1)}{2}$ generators,
leaving $d^2$ broken generators, as expected. It remains to be seen how many of those
actually couple to physical states.

In order to trigger the appeareance of a vacuum expectation value 
we have to include some dynamics to induce the
symmetry breaking. The model we propose is to add the interaction piece
\begin{equation}
  S_I = \int d^4 x (( \bar{\psi}_a \psi^{\mu} + \bar{\psi}^{\mu} \psi_a)
  B^a_{\mu} + c \det (B^a_{\mu}))
\end{equation}
Note that the interaction term also behaves as a density thanks to the
covariant Levi-Civita symbol hidden in the determinant of $B^a_\mu$.
If we consider the equation of motion for the auxiliary field $B^a_\mu$ 
we get
\be
<\bar \psi_a \psi^\mu> = 2 c \epsilon^{\mu\nu}\epsilon_{ab} B^b_\nu.
\ee
So the vacuum expectation value of the field $B$ would correspond to the 
value of the n-bein, up to a (dimensional) constant.

In what follows we shall consider the above model for  $D=2$ for simplicity.
Note the peculiar 'free' kinetic term $\gamma^a \otimes k_\mu$.
We write explicitly in two dimensions the bilinear operator
acting on the fermion fields. Note that indices $a,b,...$ can be raised and
lowered freely in Euclidean space.
\begin{equation}
     M = \left( \begin{array}{cccc}
       B_{11} &  k_1 & B_{12} &  \hspace{0.25em} k_2\\
        \hspace{0.25em} k_1 & B_{11} &  \hspace{0.25em} k_2 & B_{12}\\
       B_{21} & - i k_1 & B_{22} & - i k_2\\
       i k_1 & B_{21} & i k_2 & B_{22}
     \end{array} \right)
\end{equation}
and we also define
\be
\Delta^{ab} \equiv M M^\dagger\equiv \sum_\mu i{D}_{\mu}^{a}\cdot i{D}_{\mu}^{b},
\ee
where
\be
{D}_{\mu}^{a}= \gamma^a (\partial_\mu + i w_\mu\sigma_3) -i B_\mu^a.
\ee

We want to compute the effective action after integration of the
fermion degrees of freedom using the heat kernel method. Then
\begin{equation}
W=-\frac{1}{2}\int_{0}^{\infty}\frac{dt}{t}\text{tr}
\left<x|e^{-t\Delta}|x\right>,
\end{equation}
\begin{equation}
\begin{aligned}
\left<x|e^{-t\Delta}|x\right>=&\frac{1}{t^{D/2}}\int\frac{d^{D}k}{(2\pi)^{D}}\text{tr}\left[e^{-k^{2}\gamma^{a}\gamma^{b}+i\sqrt{t}(\gamma^{a}{D}_{\mu}^{b}k_{\mu}+{D}_{\mu}^{a}k_{\mu}\gamma^{b})+t{D}_{\mu}^{a}{D}_{\mu}^{b}}\right]\\
\end{aligned}
\end{equation}
where $\Delta$ has been defined above. Note that the exponent is a matrix in Lorentz and Dirac 
indices (the latter not explicitly written).
Once we know $W(w,B)$ we can
differentiate with respect $B^a_\mu$ and obtain the relation between the 'n-bein' and the
spin connection using a logic similar to the one defined by the Palatini formalism\cite{pa}.

Note that
\be
e^{-k^{2}\gamma^{a}\gamma^{b}}=\delta^{ab}-\frac{1}{D}\gamma^{a}\gamma^{b}+\frac{1}{D}\gamma^{a}\gamma^{b}e^{-Dk^{2}}
  \equiv P^{ab}+\frac{1}{D}\gamma^{a}\gamma^{b}e^{-Dk^{2}}
\ee
Thus the exponential, considered as a matrix, has zero modes and therefore the
heat kernel calculation is non-standard and quite laborious.

Here we shall limit ourselves to the case where there is no connection at all
and then indicate how one could proceed beyond that (rather trivial) limit, to include
a non-zero spin connection. We refer the interested reader to \cite{aep} for more details.

If $w=0$ then one can use homogeneity
and isotropy arguments to look for constant solutions of the gap equation
associated to the following effective potential
\be 
V_{eff} = c \det (B^a_{\mu}) + 2 \int \frac{d^n k}{(2 \pi)^n}
   {\rm Tr\;} (\log (- \gamma^a_{} k_{\mu} + B^a_{\mu})).
\ee
The extremum of $V_{eff}$ are found from
\be c n \epsilon_{a a_2 \ldots .a_n} \epsilon^{\mu \mu_2 \ldots . \mu_n}
   B^{a_2}_{\mu_2} \ldots .B^{a_n}_{\mu_n} + 2 {\rm tr\;} \int \frac{d^n k}{(2
   \pi)^n} (- \gamma\otimes k + B)^{- 1}\vert^\mu_a = 0.
\ee
\noindent
Notice that the equations are invariant under the permutation
\be B_{i j} \rightarrow B_{\sigma (i) \sigma (j)}, k_i \rightarrow k_{\sigma
   (i)}, \quad \sigma \epsilon S_2.
\ee
The `gap equation' to solve for constant values of $B_{i j}$ is
\be
c B_{i j} - \frac{1}{16\pi}B_{i j}\log \frac{\det B}{\mu}= 0.
\ee
A logarithmic divergence has been absorbed in $c$. This equation has 
a non-trivial solution  that we can always choose, as indicated before,
to be $B^a_\mu \sim \delta_a^\mu$.

The next step is to consider $w_\mu(x) \neq 0$. It is technically
convenient to consider the heat kernel for the operator $M^\dagger M$ rather
than $MM^\dagger$, although of course the determinants are identical. It is also
important to maintain a covariant appeareance as long as possible (note
that there is no 'metric' so far and no way of lowering or raising indices).
The final result has to be of course covariant, since our starting point is.

In conclusion, 
this leads us to the evaluation of the effective action
\begin{equation}
W=-\frac{1}{2}\int_{0}^{\infty}\frac{dt}{t}\text{tr}
\left<x|e^{-t\Delta}|x\right>
\end{equation} 
where now
\be
\Delta \equiv {\cal M}^\dagger {\cal M},
\ee
with
\begin{equation}
{\cal M}= i \mathcal{D}_{\mu}^{b},\qquad {\cal M}^{\dagger}= i \mathcal{D}_{\nu b}
\end{equation}
and
\begin{equation}
\mathcal{D}_{\mu}^{b}=\xi_{La}^{\dagger~ b}\gamma^{a}(\partial_{\rho}+iw_{\rho}\sigma_{3})\xi_{R~\mu}^{\rho}-i\bar{B}_{\mu}^{b},\,\,
\mathcal{D}_{\nu b}=\xi_{R\nu}^{\dagger~ \sigma}(\partial_{\sigma}+iw_{\sigma}\sigma_{3})\gamma_{a}\xi_{L~b}^{a}-i\bar{B}_{\nu b}.
\end{equation}
$\Delta$ now has coordinate (and Dirac) indices.
In the previous expressions we have decomposed
\begin{equation}
B_{\mu}^{a}=\xi_{L~b}^{a}\bar{B}_{\nu}^{b}\xi_{R~\mu}^{-1\nu};\quad \bar{B}_{\nu}^{b}=\xi_{La}^{\dagger ~b}B_{\mu}^{a}\xi_{R~\nu}^{\mu};\quad \bar{B}_{\nu b}=\xi_{R\nu}^{\dagger~\mu}B_{\mu a}\xi_{L~b}^{a}
\end{equation}
where $\bar{B}_{\mu}^{b}=M\delta_{\mu}^{b}$ is the backgroud which we can take to 
play the role of a mass term in the integration over $t$ in the heat kernel. Note that we have
redefined the fermion fields to absorb the matrices $\xi_L$ and $\xi_R$. 

This way of doing things ensures the formal covariance of the heat kernel expansion. It is 
not too difficult to see that the lowest non-trivial order gives
\begin{equation}
W=\frac{\mu^{2}e^{\tilde{c}}}{16\pi} ,
\int d^{2}x\sqrt{\text{Det}[(\xi_{R\,\mu}^{\sigma}\xi_{R\,\mu}^{\dagger\rho})^{-1}]},
\end{equation}
where a summation over $\mu$ is to be understood and where
$M^2=\mu^2e^{\tilde{c}}$ with $\tilde{c}=16\pi c-1$.
This is just the expected cosmological term with 
$g^{\sigma\rho}= \sum_\mu \xi_{R\,\mu}^{\sigma}\xi_{R\,\mu}^{\dagger\rho}$.

The next term in the heat kernel expansion should produce the relation ensuring that
the metric is compatible with the spin connection. Finally one would allow the spin 
connection to be a dynamical variable.

As mentioned before, there is apparently a fundamental problem in considering theories
where the graviton is generated dynamically. If we refer to the original paper by Weinberg 
and Witten\cite{ww}, the apparent pathology of these theories lies in the fact that the energy-momentum
tensor has to be identically zero if particles with spin higher than one appear. Actually,
at a very naive level the energy-momentum tensor of the toy model presented here {\it is} zero
as the model contains no metric with respect to which one can derive. Probably a energy-momentum 
tensor could be defined in some way, but this is not totally obvious, and it is not clear
to what extent the conditions assumed by Weinberg and Witten apply.

The previous two-dimensional example is all too trivial but it shows perfectly the general
ideas. It seems conceivable to entertain the idea that a mechanism analogous to chiral
symmetry breaking may trigger the dynamical appeareance of some degrees of freedom that
at the very least reproduce formally Einstein-Hilbert action. This lead to rather
interesting results, for instance we expect the following relation between the Planck mass
and the dynamically generated mass
\be
M_{P}^2\sim \frac{M^2}{16\pi^2}\log \frac{\mu}{M}.
\ee
We have also seen above how a relation between the would-be cosmological constant
and the parameters of the underlying theory appears. 

This is probably an appropriate place to stop and we recommend to the interested reader 
to examine the results that will be presented in \cite{aep}.

\section{Summary}

In these lectures we review the physical consequences of treating gravity at 
the quantum level as an effective theory, not very different from what is done 
in pion physics. Because it contains massless states, non-local logarithmic terms
in the effective action should then be present.

We have analyzed the relevance of the non-local
quantum corrections due to the virtual exchange of gravitons and other
massless modes to the 
evolution of the cosmological scale factor in FRW universes. 
The effect is largest in a de Sitter universe with a large
cosmological constant. The effects are nonetheless locally absolutely tiny, but they lead 
to a noticeable secular effect that slows down the inflationay expansion. 
Although this has not been discussed in detail in these lectures,
in a matter dominated universe the effect
is a lot smaller, and it appears to be of the opposite sign. Quantum 
effects seem to enhance the expansion rate in this case.
These effects have no classical
analogy.

Note that the results presented
here are not `just another model'. Quantum gravity non-local 
loop corrections exist. They are required by unitarity if gravity is to be
a consistent quantum theory. The non-localities also give rise to other consequences; 
for instance it would be very interesting to compute the space correlations that these
logarithmic terms introduce.

In the final part we have discussed 
a toy model where gravitons appear as a Goldstone states. 
The model has originally no metric whatsoever; it is generated dynamically.

\section*{Acknowledgements}
We acknowledge the financial support from the RTN ENRAGE and the reserch projects 
FPA2007-66665 and SGR2009SGR502.
We thank
A. A. Andrianov for disccussions on the subject. These notes were finalized at the PH Department 
at CERN. Finally, it is a pleasure to thank the
organizers of the Zakopane School on Theoretical Physics for the excellent organization
and warm hospitality.

\end{document}